\documentclass[twocolumn, aps, prx, reprint]{revtex4-2}
\usepackage{physics}
\usepackage{microtype}
\usepackage{graphicx}
\usepackage{svg}
\usepackage{amsmath, amssymb}
\usepackage{dsfont}
\usepackage{booktabs}
\usepackage{hyperref}
\usepackage{caption, subcaption}
\usepackage{xspace}
\usepackage{float}
\usepackage{algorithm}
\usepackage{algpseudocode}

\newcommand{\diff}{\textsc{Diff}}
\newcommand{\fd}{\textsc{FD}}

\DeclareMathOperator{\pr}{Pr}
\DeclareMathOperator*{\expec}{\mathds{E}}
\DeclareMathOperator{\variance}{Var}

\renewcommand{\pl}{{\ensuremath{p_L}\xspace}}

\newcommand{\pvec}{\mathbf p}

\begin{document}

\title{Efficient Estimation of Logical Sensitivities Through Fault-Counting}
\author{Winston Fu\textsuperscript{1}}
\email{winston.fu@princeton.edu}
\author{J. Wilson Staples\textsuperscript{1}}
\author{Jeff D. Thompson\textsuperscript{1,2}}
\affiliation{\textsuperscript{1}Princeton Quantum Initiative, Princeton University, Princeton, NJ 08544, USA}
\affiliation{\textsuperscript{2}Department of Electrical and Computer Engineering, Princeton University, Princeton, NJ 08544, USA}

\noindent

\date{\today}

\begin{abstract}

Quantum error-correcting circuits are affected by multiple physical noise mechanisms, whose contributions to logical failure must be understood to evaluate code performance and guide improvements in hardware.
The individual error budget contributions for an error type $i$ can be characterized by its logical sensitivity $\nu_i = \pdv*{p_L}{p_i}$, which measures the response of the logical error rate $p_L$ to each physical noise parameter $p_i$. Normally, $\nu_i$ is measured using linear fits such as finite differences, where $p_L$ is measured at two or more values of $p_i$ to calculate partial derivatives. In this paper, we develop a differentiable estimator to obtain all components of $\nabla_{\mathbf p}p_L$ simultaneously from a single Monte Carlo data set at one noise configuration, using information about the underlying fault configurations.
In surface code simulations with circuit-level noise, the estimator agrees with conventional finite differences while requiring one to two orders of magnitude fewer shots to achieve the same variance. We apply this technique to quantify error budgets, resolve sensitivities at the individual qubit level, and infer effective code distance. Finally, we incorporate the sensitivities into Newton root finding to locate and trace threshold contours in multidimensional noise models. These results provide an efficient method for extracting and applying logical sensitivity information from standard quantum error correction simulations.

\end{abstract}

\maketitle

\section{Introduction}
Quantum error correction (QEC) suppresses physical noise by encoding information redundantly and non-locally. 
Physical qubits are subject to multiple noise mechanisms, and understanding how each component contributes to logical failure is important for accurately estimating code performance \cite{iyerSmallQuantumComputer2017,chenExponentialSuppressionBit2021,acharyaSuppressingQuantumErrors2023,acharyaQuantumErrorCorrection2025,deMartiiOliusPerformanceSurfaceCodes2022,puttermanHardwareefficientQuantumError2025a}. Real devices contain distinct error mechanisms associated with gates, idling, state preparation, and measurement, whose relative importance can differ substantially.

A natural local characterization of this dependence is the logical sensitivity, $\nu_i(\pvec)\equiv \pdv*{p_L(\pvec)}{p_i}$, where $\pvec=(p_1,\ldots,p_s)$
is the vector of physical noise parameters and $p_i$ is the physical error rate associated with noise channel $i$. Reducing the error rate of an error type with higher logical sensitivity would cause a larger reduction in logical error rate. The dimension $s$ of $\pvec$ depends on the features of the error model that are the subject of study, and can range from simply grouping errors into types (single-qubit, two-qubit, idle, etc.) to parametrizing individual space-time locations~\cite{chadwickErasureMinesweeperExploring2025,carrollSubsystemSurfaceCompass2024,tiurevCorrectingNonindependentNonidentically2023}.

The standard numerical approach estimates each $\nu_i$ by finite differences \cite{chenExponentialSuppressionBit2021,acharyaSuppressingQuantumErrors2023,acharyaQuantumErrorCorrection2025,eickbuschDemonstrationDynamicSurface2025a,puttermanHardwareefficientQuantumError2025a}: perturb $p_i$, rerun the circuit simulation and decoder, and divide the resulting change in $p_L$ by the perturbation size. This incurs a separate set of simulations for every parameter, so its cost grows linearly with $s$. The step size must be small enough to control Taylor series truncation error, but large enough for the change in $p_L$ to be statistically resolved. These limitations lead to a high computational cost when characterizing circuit-level noise models with fine-grained parameterization of $\pvec$.

In experiments on quantum devices, we have access to the measured syndromes and the logical outcome of the decoder. However, in stabilizer tableau simulations, we also have access to the underlying fault configurations that produced these syndromes. While usually treated as internal bookkeeping, these fault configurations contain additional information that can be used to efficiently estimate quantities other than the logical error rate.

\begin{figure*}
    \centering
    \includegraphics[scale=1]{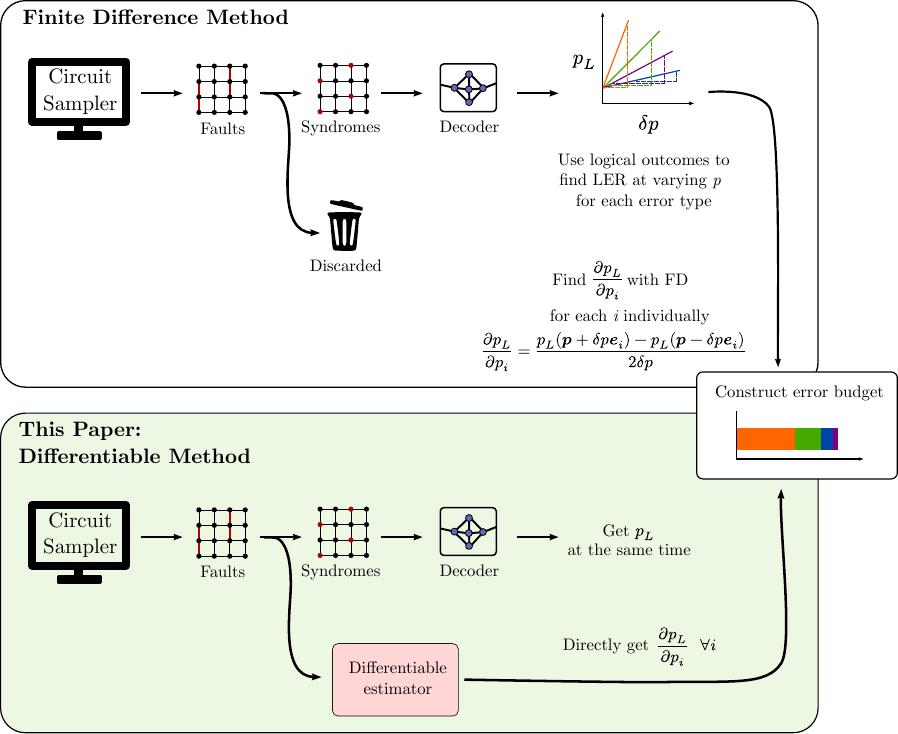}
    \caption{Overview of contributions. Top: The standard approach to finding sensitivities and error budgets is the finite difference method. This involves using values of $p_L$ measured at various $p$ to calculate gradients, requiring multiple simulations at different $\vb p$. $\vb e_i$ is the unit vector in the $p_i$ direction. Bottom: We introduce a differentiable estimator to estimate sensitivities using samples at only one $\vb p$. The sensitivity to different components of $\vb p$ is extracted from the sampled fault locations, which are ordinarily discarded if the only object is to compute $p_L$.
    }
    \label{fig:overview}
\end{figure*}

In this paper, we develop a differentiable estimator that obtains all components of $\nabla_{\pvec}p_L$ from a single Monte Carlo data set (Fig.~\ref{fig:overview}). It is a novel application of the score-function estimator \cite{rubinsteinScoreFunctionApproach1986,glynnLikelihoodRatioGradient1990}, which has previously been used in QEC for reinforcement-learning-based optimization of decoder priors \cite{sivakOptimizationDecoderPriors2024}, control parameters \cite{sivakRealtimeQuantumError2023,sivakReinforcementLearningControl2026}, and code constructions \cite{olleSimultaneousDiscoveryQuantum2024,suDiscoveryOptimalQuantum2025}, where the score function is taken with respect to a distribution used to sample candidate parameter choices. Instead, here we apply the score function directly to the distribution of physical fault configurations, yielding derivatives of the logical error rate with respect to the underlying physical noise parameters from the same samples used to estimate $p_L$.

We validate the estimator in circuit-level surface code simulations by comparing it directly with finite differences across physical error rates and code distances. We then test this method as a subroutine in an algorithm which maps threshold surfaces in multi-parameter noise models \cite{dawsonNoiseThresholdsOptical2006,svoreFlowmapModelAnalyzing2006,staceThresholdsTopologicalCodes2009,bonillaAtaidesXZZXSurfaceCode2021,vodolaFundamentalThresholds2022}. These results establish logical sensitivity as an easily accessible diagnostic for QEC design and hardware optimization that can be estimated from the same simulation samples used to estimate logical error rates.

\section{Efficient Sensitivity Estimation}
\subsection{Algorithm}

Consider a stochastic error model consisting of a set $\Omega$ of independently sampled error mechanisms. The model is governed by a set of physical error parameters $\vb p$. The parametrization may use a single shared error rate, as is the case in a code-capacity noise model, or every spacetime fault location $f \in \Phi$ may be assigned a separate parameter $p_f$. The commonly used circuit-level noise model is an intermediate example, where faults of the same error type use a shared parameter. Each error mechanism $e\in\Omega$ is sampled with probability $W_e(\pvec)$. For a given shot, let $E\subseteq\Omega$ denote the set of mechanisms that occur, and let the decoder outcome be represented by the indicator function
\begin{equation}
        L(E)
    =
    \mathds{1}\{E \text{ causes a logical error}\},
    \qquad
    E\subseteq\Omega.
\end{equation}
The logical error rate can be written as
\begin{equation}
    \pl(\pvec)
    =
    \sum_{E \subseteq \Omega}
    L(E)\,\pr_{\pvec}(E) = \expec_{E \sim \pr_{\vb p}}[L(E)],
\end{equation}
where, under the assumption that faults are sampled independently,
\begin{equation}
    \pr_{\pvec}(E)
    =
    \prod_{e \in E} W_e(\pvec)
    \prod_{e \notin E} \bigl(1-W_e(\pvec)\bigr).
\end{equation}

The logical sensitivity can then be expressed as:
\begin{align}
    \nu_i(\pvec)
    &\equiv
    \pdv{\pl(\pvec)}{p_i}
    \notag\\
    &=
    \sum_{E\subseteq\Omega}
    L(E)\,\pr_{\pvec}(E)
    \pdv{p_i}\log \pr_{\pvec}(E)
    \notag\\
    &=
    \expec_{E\sim\pr_{\pvec}}
    \left[
        L(E)
        \pdv{p_i}\log \pr_{\pvec}(E)
    \right]
    \notag\\
    &=
    \expec_{E\sim\pr_{\pvec}}
    \Biggl[
        L(E)\Biggl(
            \sum_{e\in E}
            \frac{1}{W_e(\pvec)}
            \pdv{W_e(\pvec)}{p_i}
    \notag\\
    &\qquad\qquad {}
            -
            \sum_{e\notin E}
            \frac{1}{1-W_e(\pvec)}
            \pdv{W_e(\pvec)}{p_i}
        \Biggr)
    \Biggr].
    \label{eq:sens_model}
\end{align}
Intuitively, the sensitivity measures how strongly faults of a given type are represented in failing fault configurations. Every sensitivity $\nu_i(\vb p)$ can be estimated from the same set of data that was used to estimate $p_L(\vb p)$. Additionally, under the fault-level parametrization, the fact that all $\nu_f$ can be found at once means that the grouping of faults does not have to be decided beforehand. Instead, the sensitivity for any grouping of faults $\nu_{F}$ can be found by storing and summing the individual sensitivity components $\nu_f$. While second- and higher-order derivatives can also be calculated using this method, we will consider only first-order derivatives in this paper for simplicity.

\subsection{Numerics}

We benchmarked the differentiable method by running simulations of the surface code with circuit-level noise in Stim \cite{gidneyStimFastStabilizer2021}. Details of how the sampler calculates $W_e(\vb p)$ given $\vb p$ are given in Appendix~\ref{app:inc_err}.

We compared the differentiable estimator (\diff{}) with the 
finite-difference estimator (\fd{}), corresponding to the simplest local straight-line fit. The differentiable estimator given $N$ shots from Eq.~\ref{eq:sens_model} is
\begin{equation}
    \hat{\nu}_i^{\mathrm{Diff}}
    =
    \frac{1}{N}
    \sum_{k=1}^{N}
    L(E_k)
    \pdv{p_i}
    \log \pr_{\pvec}(E_k).
\end{equation}

The finite-difference estimator is:
\begin{equation}
    \hat{\nu}_i^{\mathrm{FD}}
    =
    \frac{
        \hat p_{L, d}(\pvec + h \mathbf e_i)
        -
        \hat p_{L, d}(\pvec - h \mathbf e_i)
    }{2h},
\end{equation}
where $\hat p_{L,d}(\vb p) = \frac{1}{N} \sum_{k=1}^{N} L(E_k)$ with error mechanisms $E_k$ sampled from the distribution corresponding to the distance-$d$ circuit parametrized by $\pvec$. $h$ is the finite-difference step size, and $\mathbf e_i$ denotes the unit vector in the
$p_i$ direction. The decoder is held fixed at $\vb p$.

\begin{figure*}
    \centering
    \includegraphics[scale=1]{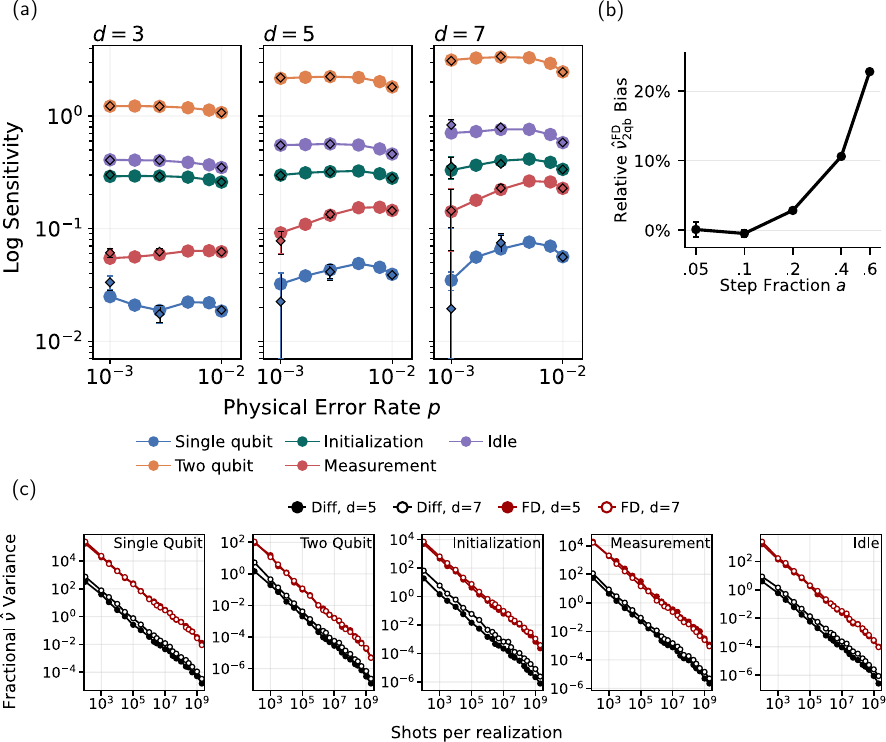}
    \caption{(a) Logarithmic sensitivities $\pdv*{\log p_L}{\log p_i}$, over a range of $p$. Circles are the estimates obtained using \diff{}, while diamonds are estimates obtained using \fd{}. (b) Relative difference between $\hat \nu_{2\text{qb}}^\text{FD}$ and $\hat \nu_{2\text{qb}}^\text{Diff}$ at different step fractions $a$, for $d=7$ at $p=0.005$. As the sensitivity is measured over a larger range of $p$, higher-order terms cause a bias in the measured sensitivity. (c) Fractional variance in sensitivity from \diff{} and \fd{} at varying shot budgets, obtained from 200 bootstrap samples of $2\times 10^9$ shots for each error type at step fraction $a=0.1$.}
    \label{fig:diff_vs_fd}
\end{figure*}

We first tested the accuracy of \diff{} over a range of $10^{-3} \leq p \leq 10^{-2}$, $p_i = p$ for all $i$, under circuit-level noise for the unrotated surface code for distances 3, 5, and 7. \diff{} was used to estimate the individual sensitivities $\nu_i$ at 6 points with $10^9$ shots each. These were compared at three $p$ values with \fd{} with step fraction $a = h/p_i = 0.1$. For \fd{}, we ran two simulations per error type, with $10^9$ shots
per simulation except at $p=10^{-3}$, where we used $3\times10^9$. For a clearer comparison across error scales, we compared the logarithmic sensitivity, $\pdv*{\log p_L}{\log p_i} = \nu_i p_i/p_L$, in Fig.~\ref{fig:diff_vs_fd}a. The small values of $\nu_i$ for single-qubit gate and measurement errors made it difficult to measure precisely without introducing bias with \fd{}, leading to large error bars at $p=10^{-3}$. The two estimators displayed close numerical agreement across the configurations tested.

The step fraction was chosen to limit finite-step bias in \fd{}. For example, Fig.~\ref{fig:diff_vs_fd}b shows the difference between $\hat \nu_{2\text{qb}}^\text{FD}$ and $\hat \nu_{2\text{qb}}^\text{Diff}$ as $a$ was varied at $d=7$, $p=0.005$. A measurable bias was observed at $a=0.2$, and continued increasing with larger step sizes.

Next, we compared the performance of the two estimators when calculating $\nu_i$ for 5 error types under equivalent shot budgets (Fig.~\ref{fig:diff_vs_fd}c). Using $a=0.1$, we measured the variance when using a fixed number of shots to measure $\nu$ for 5 error types at $p=0.005$. This was done over a range of shot budgets between 100 and $2\times 10^9$, where the data for smaller shot budgets were obtained through 200 bootstrap resamples of the data set collected at the largest shot budget. The variance ratios between the two estimators remained approximately constant across the different shot budgets. The values agree with the predicted ratio that we expect from treating logical failures as Bernoulli random events (Appendix~\ref{app:uncertainty}):
\begin{equation}
    R_i = \frac{\variance(\hat{\nu}_i^{\mathrm{FD}})}{\variance(\hat{\nu}_i^{\mathrm{Diff}})}
    \approx
    \frac{s}{a^2 \mathbb E[M_i^2 \mid L]},
    \label{eq:variance_ratio}
\end{equation}
where $\mathbb E[M_i^2 \mid L]$ is the expected squared number of channel-$i$ faults conditioned on logical errors.  The factor of $s$ in Eq.~\ref{eq:variance_ratio} accounts for the splitting of shot budget using each estimator. Using \diff{}, one could use the entire shot budget on a single experiment at $p=0.005$. However, using \fd{}, one needed to split the shot budget over 10 separate experiments: two for each of the 5 error types at $p_i = (1 \pm 0.1)p$. Furthermore, sampling noise is amplified by the size of the finite step in \fd{}. Hence, for the configurations tested, \diff{} reduces the number of shots required to estimate $\nu_i$ by one to two orders of magnitude over \fd{}.

\begin{figure*}
    \centering
    \includegraphics[scale=1]{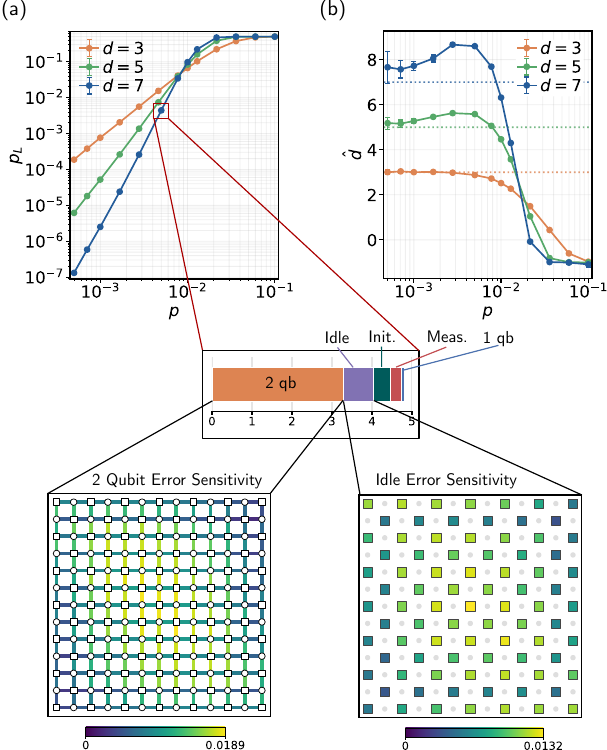}
    \caption{(a) A $p_L$ vs $p$ sweep of a surface code at distances 3, 5, and 7. Using \diff{}, the samples collected at each point on the curve can also be used to estimate a breakdown of $\nu_i$ by error type, round number, spatial location, and individual fault. The spatial sensitivity heatmaps, aggregated over $d$ rounds, are shown for two-qubit gate errors and idle errors, where squares represent data qubits and circles represent ancilla qubits. The plotted values are logarithmic sensitivities, $\pdv*{\log p_L}{\log p_i} $. (b) At the same time, the sum of sensitivities gives a local unbiased measurement of the slope of the $p_L$ vs $p$ curve, which can be used to calculate the local effective distance scaling factor. In the low-$p$ limit, this value corresponds to the code distance (dotted lines), while in the high-$p$ limit, all distances converge to $-1$ as $p_L$ plateaus at 1/2.}
    \label{fig:diff_resolution}
\end{figure*}

\diff{} allows for the measurement of sensitivities in a circuit at a per-fault level, using the same simulation data used to estimate $p_L$ with additional processing of the sampled fault configurations. For our noise model, this corresponds to $s=268,\ 1526$, and 4544 for distances 3, 5, and 7 respectively. $\mathbb E[M_i^2 \mid L]$ in Eq.~\ref{eq:variance_ratio} scales at least as $1/s$ when subdividing a fixed number of faults, hence $R_i$ scales with the square of the number of error types defined for the circuit. Fig.~\ref{fig:diff_resolution}a shows the varying levels of resolution possible when estimating sensitivities with \diff{}. Firstly, along every point of a $p_L$ vs $p$ curve, the sensitivities can be aggregated by error type. Such decompositions help identify which physical processes to prioritize for improvement, as illustrated by experimental error budgets separating gate, measurement, idling, and qubit loss contributions~\cite{acharyaQuantumErrorCorrection2025,eickbuschDemonstrationDynamicSurface2025a,puttermanHardwareefficientQuantumError2025a}. These values at different distances can also be used to calculate logical error scaling budget components:
\begin{equation}
    \hat{\pdv{\Lambda_d^{-1}}{p_i}} =  \frac{\hat p_{L,d+2}}{\hat p_{L,d}} \left(
    \frac{\hat{\nu}^{\mathrm{Diff}}_{d+2,i}}{\hat p_{L,d+2}}
    - \frac{\hat{\nu}^{\mathrm{Diff}}_{d,i}}{\hat p_{L,d}}
  \right),
  \label{eq:lambda_comp}
\end{equation}
where $\Lambda_d = p_{L, d} / p_{L, d+2}$.

However, recent studies in hybrid-erasure architectures \cite{chadwickErasureMinesweeperExploring2025} and heterogeneous noise models \cite{deMartiiOliusPerformanceSurfaceCodes2022,Khosravani:2026jma} have revealed a strong spatial dependence of the logical performance of error-correcting codes on individual qubit characteristics. Motivated by these examples, we demonstrate applying \diff{} to resolve the position-dependent sensitivity to each error type. For instance, the plotted spatial heatmaps show a higher relative importance of qubits in the bulk compared to the boundary, with sensitivities decreasing from the center of the code patch. These heatmaps can help inform decisions about resource allocation for hardware improvements. Thus, a single simulation campaign supports both coarse hardware error budgets and fine-grained circuit diagnostics, with the grouping chosen after sampling rather than fixed in advance. 

\diff{} can also be used to find the local effective distance scaling factor at every point. Along the uniform-noise path $p_i=p$ for all $i$, define $\hat\nu^{\mathrm{Diff}} = \sum_i \hat\nu^{\mathrm{Diff}}_i$. This estimates the total derivative $\dv*{p_L}{p}$. Starting from the ansatz $\pl = A\left( {p}/{p_\text{th}} \right)^{(d+1)/2 }$, we differentiate to obtain the distance estimator $\hat{d} = 2\left( \hat\nu^{\mathrm{Diff}} {p}/{\hat p_L} \right) - 1$. This is shown in Fig.~\ref{fig:diff_resolution}b, and provides a useful diagnostic of fault-tolerance properties of a circuit.

\newpage
\section{Efficient Threshold Estimation}

\subsection{Algorithm}

Finally, we applied our efficient sensitivity estimator to the problem of finding finite-distance code thresholds. While the threshold error rate of a code is often taken to be a single scalar value, a more accurate picture is to consider a surface in the multidimensional error space.
Such threshold surfaces have been studied for a variety of codes and noise models, including concatenated codes with multiple circuit-level error rates \cite{svoreFlowmapModelAnalyzing2006,dawsonNoiseThresholdsOptical2006}, surface codes subject to qubit loss \cite{staceThresholdsTopologicalCodes2009}, and surface codes under biased Pauli noise \cite{bonillaAtaidesXZZXSurfaceCode2021}.
The usual method for finding the threshold of a QEC code involves sampling at a range of error rates close to the threshold, then fitting the data to a model \cite{svoreFlowmapModelAnalyzing2006,staceThresholdsTopologicalCodes2009,bonillaAtaidesXZZXSurfaceCode2021,brownFaulttolerantErrorCorrection2016}. Doing so is computationally expensive, as it requires sampling over a wide range of error parameters. This problem is made worse when finding the threshold surface for multiple error types. If $\mathcal{O}(K)$ points were needed to find the threshold for an error rate parametrized by a scalar, then $s$ error types would require repeating this process many times. Up to $\mathcal{O}(K^s)$ points need to be sampled to fully cover the threshold surface. This has motivated the use of approximate threshold-finding algorithms, such as detecting crossings in logical error rate from samples at varying distances and error rates \cite{zenLowvalencyScalableQuantum2026}.
Furthermore, it has been shown that the critical exponent model used for threshold fitting does not account for all finite-size corrections \cite{xiaoExactResultsFinite2024}, making it desirable to estimate the threshold in a model-agnostic manner.

The algorithm we propose employs sensitivity information in Newton--Raphson root finding to locate the crossing in logical error rate between two code distances. Define the crossing function:
\begin{equation}
    \Delta(p) \equiv \Delta_{d_1, d_2}(p) = \log p_L(d_2, p) - \log p_L(d_1, p).
\end{equation}
The root of this equation, $\Delta(p_\text{th}) = 0$, gives the finite-distance threshold with respect to the two distances $d_1$ and $d_2$.
Then, we can iteratively find the root with the standard procedure:
\begin{equation}
    p^{(1)} = p^{(0)} - \frac{\Delta(p^{(0)})}{\Delta'(p^{(0)})},
\end{equation}
where
\begin{equation}
    \Delta'(p) = \frac{1}{p_L(d_2, p)}  \nu(d_2, p) - \frac{1}{p_L(d_1, p)} \nu(d_1, p).
\end{equation}

In the $s$-dimensional case, $\Delta$ is a function of the full physical-noise vector,
\begin{equation}
    \Delta(\pvec)
    =
    \log p_L(d_2,\pvec)-\log p_L(d_1,\pvec),
\end{equation}
\begin{equation}
    \pvec^{(1)}
    =
    \pvec^{(0)}
    -
    \frac{\nabla_{\pvec}\Delta(\pvec^{(0)})}
    {\left\|\nabla_{\pvec}\Delta(\pvec^{(0)})\right\|^2}
    \Delta(\pvec^{(0)}),
\end{equation}
where $\pvec^{(0)}$ and $\pvec^{(1)}$ are successive points in the $s$-dimensional physical-noise space.

By using $\hat\nu^{\mathrm{Diff}}$ as the estimator for $\nu$, a pair of experiments, one at each distance, is sufficient to measure both $\Delta(\pvec)$ and $\nabla_{\pvec}\Delta(\pvec)$ simultaneously.

After finding one point on the threshold hypersurface, we also learn its normal vector, giving us $(s-1)$ linearly independent tangent directions in probability space where $\Delta$ remains constant to first order. In the $s=2$ case, this corresponds to a local tangent to the threshold contour line. Note that as we are working at finite distances, the resulting crossing contours only approximate the asymptotic threshold.

\begin{figure*}
    \centering
    \includegraphics[scale=1]{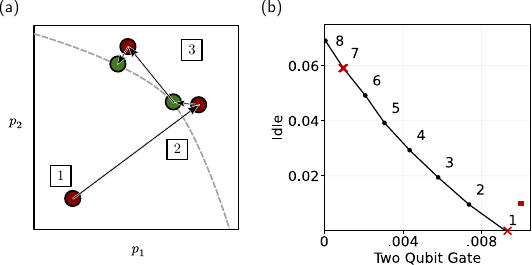}
    \caption{(a) Illustration of the threshold-finding algorithm for the case $s=2$. The gray line shows the threshold contour between error types $p_1$ and $p_2$ that needs to be found. Boxed numbers represent the steps of the algorithm. Step 1: For two distances $d_1$ and $d_2$, sample the circuits at some $\vb p_0$, measure $p_L$, and calculate $\Delta(\vb p_0)$, while also using \diff{} to find $\grad_{\vb p} \Delta(\vb p_0)$ from the same set of data. Step 2: The Newton step predicts a new point $\vb p_1$ for which $\Delta(\vb p_1) = 0$. Repeat until the convergence criteria are met. Step 3: Choosing a $\vb p_2$ where $\vb p_2 - \vb p_1$ is orthogonal to the vector $\grad_{\vb p} \Delta(\vb p_1)$. This leaves $\Delta(\vb p_2) = 0$ to first order, and so is close to the threshold as well. Repeat root finding until this next point converges, and continue until the desired range of the threshold contour is obtained. (b) Data from a simulation of the surface code at distances 5 and 7, decoded with PyMatching, that shows all the points that needed to be sampled to trace the threshold contour between two-qubit gate and idle errors. Red square: initial point. Red crosses: points that did not fit the convergence criteria, and required further Newton steps. Black circles: converged points that lie on the crossing between $d=5$ and 7 codes.}
    \label{fig:thres_finder_explanation}
\end{figure*}

For the two-dimensional slices considered below, the complete contour-tracing procedure is illustrated in Fig.~\ref{fig:thres_finder_explanation}a and detailed in Algorithm~\ref{alg:threshold_contour}.

\begin{algorithm}[H]
\caption{Sensitivity-guided threshold contour tracing}
\label{alg:threshold_contour}
\begin{algorithmic}[1]
\Require Initial guess $\pvec_0$, distances $d_1,d_2$, step size $h$,
         fractional tolerance $\epsilon$, number of steps $r$
\Ensure List of contour points $\mathcal{C}$

\Function{FindThres}{$\pvec$}
    \State $\mathcal D \gets \varnothing$
    \While{true}
        \State Sample shots at $\pvec$ for both distances;
        \State Append new samples to $\mathcal D$
        \State Estimate $\Delta$, $\sigma_\Delta$, and
               $\mathbf n=\nabla_{\pvec}\Delta$ from $\mathcal D$
        \If{$|\Delta|>2\sigma_\Delta$}
            \State $\displaystyle
                \pvec\gets\pvec
                -\frac{\Delta}{\|\mathbf n\|_2^2}\mathbf n$
            \State $\mathcal D\gets\varnothing$
        \ElsIf{$\displaystyle
            \frac{\sigma_\Delta}{|\pvec\cdot\mathbf n|}
            \le\epsilon$}
            \State \Return $(\pvec,\mathbf n)$
        \EndIf
    \EndWhile
\EndFunction

\State $(\pvec,\mathbf n)\gets\Call{FindThres}{\pvec_0}$
\State $\mathcal{C}\gets[\pvec]$

\For{$k=1,\ldots,r$}
    \State Choose $\mathbf t$ such that
           $\mathbf n\cdot\mathbf t=0$ and $\|\mathbf t\|_2=1$
           \Comment{tangent at the current anchor}
    \State $\pvec_{\mathrm{pred}}\gets\pvec+h\mathbf t$
    \State $(\pvec,\mathbf n)
           \gets\Call{FindThres}{\pvec_{\mathrm{pred}}}$
           \Comment{update the anchor and normal}
    \State Append $\pvec$ to $\mathcal{C}$
\EndFor

\State \Return $\mathcal{C}$
\end{algorithmic}
\end{algorithm}

\subsection{Numerics}
\begin{figure*}
    \centering
    \includegraphics[scale=1]{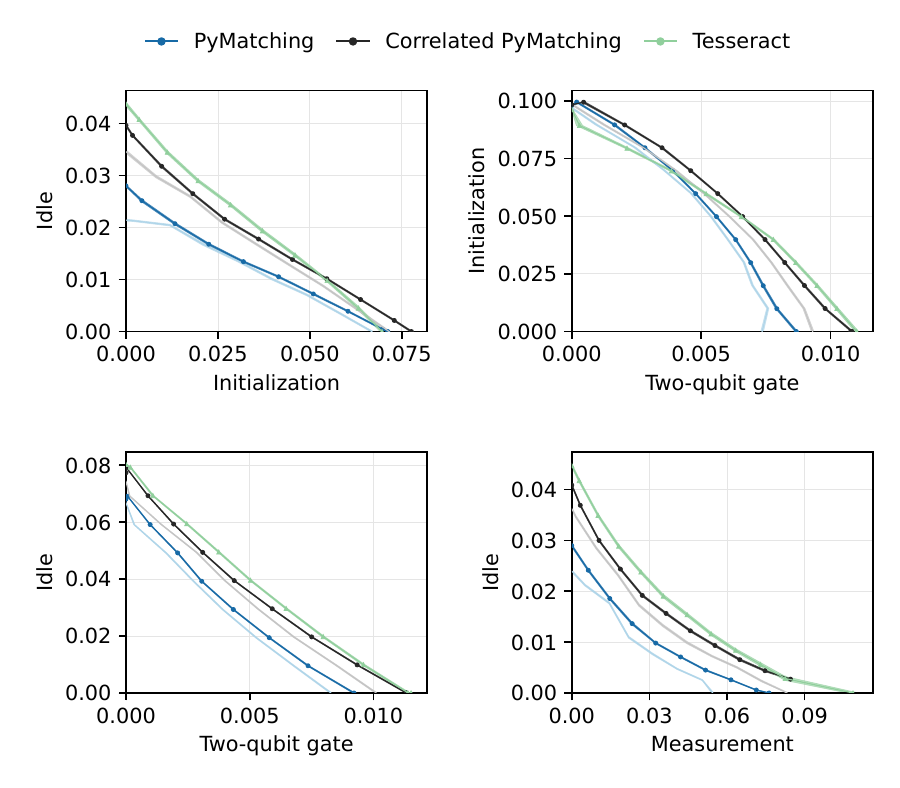}
    \caption{Threshold contours for four error type pairs, using PyMatching, Correlated PyMatching, and Tesseract. Uncertainty in contour position due to statistical uncertainty in $\Delta$ is plotted as a lightly shaded region around each contour line. For PyMatching and Correlated PyMatching, darker shades show the thresholds found using the $(5, 7)$ distance pair. Lighter shades show the $(3, 5)$ distance pair. For Tesseract, only the $(3, 5)$ distance pair was measured.}
    \label{fig:contour}
\end{figure*}

We demonstrated the effectiveness of using logical sensitivities to guide the tracing of threshold contours by running simulations on the surface code with a variety of decoders.
We ran Alg.~\ref{alg:threshold_contour} to trace the contours for the four combinations of error types, while holding the remaining error rates fixed.
We used an iterative refinement approach. First, $2\times 10^6$ shots were taken at one point. If $|\Delta| > 2 \sigma_\Delta$, then the noise parameters were updated according to Newton root finding. If $|\Delta| \leq 2 \sigma_\Delta$, then we checked whether $\sigma_\Delta/ {|\pvec\cdot\mathbf n|} \le\epsilon$, with more shots taken at that same point until this cutoff was met. This corresponds to the uncertainty in the normal direction to the contour, normalized by the perpendicular distance from the tangent to the origin. As $\sigma_\Delta$ was decreased, the residual could become statistically significant again, requiring another Newton step. Otherwise, a point was considered converged once both criteria were met.
This was done using distances $(3, 5)$ and $(5, 7)$ with PyMatching \cite{higgottSparseBlossomCorrecting2025a} and Correlated PyMatching, and $(3, 5)$ with Tesseract \cite{beniTesseractSearchBasedDecoder2025b}. Fig.~\ref{fig:thres_finder_explanation}b shows a representative run for distances (5,7) with PyMatching. All the measured contours are shown in Fig.~\ref{fig:contour}, with independent samples at each converged point verifying that all points fall within $2 \sigma$ of $\Delta=0$.

Using the sensitivity at previous points to predict subsequent points on the threshold contour reduces the number of points that must be sampled to find a crossing between codes of different distances. Fig.~\ref{fig:thres_finder_explanation}b shows the points that were sampled to trace the threshold contour between two-qubit gate and idle errors. The predicted locations
for the points where $\Delta=0$, based on local sensitivity information, lay close to the actual threshold, and required only small refinements to converge. By avoiding repeated scans across the crossing contour, the number of points needed using Newton root finding is $\mathcal{O}(K^{s-1})$, representing a factor $K$ improvement over grid sampling.

\section{Discussion}

We have presented a differentiable method that estimates logical sensitivities to all physical noise parameters from a single Monte Carlo data set. In circuit-level surface code simulations, the estimator agrees with finite differences and achieves lower variance at a fixed total shot budget. \diff{} makes use of sampled fault configurations, in combination with decoder logical outcomes, to estimate sensitivities at the individual fault level. This enables arbitrary post-hoc aggregation of sensitivities and error budgets at different levels of resolution, from individual qubits to error types.

Incorporating \diff{} into Newton root finding enables efficient threshold contour estimation in multi-parameter noise models, using model-agnostic explicit distance crossings. Together, these results provide a practical tool for diagnosing QEC circuits and identifying which physical errors most strongly affect logical performance.

For future work, the differentiable estimator could be extended to second- and higher-order derivatives. These derivatives account for the curvature of the threshold contour, allowing more accurate predictions of the threshold. In addition, this information may be applied to rare-event sampling to increase the probability of sampling harmful correlated fault configurations. Although our numerical demonstrations used Pauli noise, \diff{} applies equally to any independently sampled stochastic event, such as leakage, erasure \cite{wuErasureConversionFaulttolerant2022a,chadwickErasureMinesweeperExploring2025}, high-weight correlated errors \cite{mcewenResolvingCatastrophicError2022}, and non-Clifford errors \cite{harperNonCliffordCrosstalkNoise2026,cercelescuPolynomialtimeSimulationNonClifford2026}. These mechanisms enable more realistic modelling of quantum hardware, but can be costly to simulate. \diff{} could be used to efficiently extract the sensitivities for these noise mechanisms from simulations to better guide hardware improvements.

\textit{Note added} --- While finalizing this manuscript, we became aware of a similar approach to logical sensitivity estimation in Ref.~\cite{shaoOptimizingQuantumError2026}.

\textit{Acknowledgements} ---  We acknowledge D.E. Koh, S.T. Goh, and J.F. Kong for helpful discussions. This work was supported by the National Science Foundation (QLCI grant OMA-2120757, and PHY-2047620), the Army Research Office (W911NF-1810215), the Gordon and Betty Moore Foundation (grant DOI 10.37807/gbmf12253), the Office of Naval Research (N00014-20-1-2426), and the Sloan Foundation. Winston Fu is supported by the A*STAR National Science Scholarship.

\textit{AI Disclosure} --- The authors acknowledge the use of LLMs to assist in ideation, coding, data analysis, and proofreading the manuscript.

\textit{Competing interests} --- J.D.T. is a shareholder in Logiqal, Inc.

\bibliography{ref}

\onecolumngrid
\appendix
\section{Inclusive Error Model}
\label{app:inc_err}

In our simulations, we parametrized $\vb p$ by error types $t(e)$ for error mechanism $e$, as shown in Table~\ref{tab:fault_type_channel_class}.
Each error $e$ is sampled according to an inclusive error model \cite{chaoOptimizationSurfaceCode2020}:
\begin{equation}
    W_e(\pvec) = 
    \begin{cases}
        p_{t(e)}
        & \text{if } n_{c(e)} = 1, \\
        \dfrac12
        -
        \dfrac12
        \left(
            1-
            \dfrac{2^{n_{c(e)}}}{2^{n_{c(e)}}-1}
            p_{t(e)}
        \right)^{2^{1 - n_{c(e)}}}
        & \text{if } n_{c(e)} \geq 2.
    \end{cases},
    \label{eq:edge_weight}
\end{equation}

Each error mechanism is also assigned an error channel class $c(e)$, with a corresponding integer value for $n_{c(e)}$ (Table~\ref{tab:fault_type_channel_class}). $2^{n_{c(e)}}-1$ is the number of independent mechanisms associated with a channel $c(e)$, and $p_{t(e)}$ is the physical error rate assigned to $t(e)$.

\begin{table}
\centering
\caption{Error types $t(e)$, their channel classes $c(e)$, and the corresponding values of $n_{c(e)}$.}
\label{tab:fault_type_channel_class}
\begin{tabular}{lll}
\toprule
Channel class $c(e)$ & Example error types $t(e)$ & $n_{c(e)}$ \\
\midrule
Single-qubit bit-flip     & Measurement, initialization       & 1 \\
Single-qubit depolarizing & Single-qubit gate, idle           & 2 \\
Two-qubit depolarizing    & Two-qubit gate                    & 4 \\
\bottomrule
\end{tabular}
\end{table}

Hence,
\begin{equation}
    \pdv{W_e(\pvec)}{p_i}
    =
    \mathds{1}[t(e)=i]
    \begin{cases}
        1
        & \text{if } c(e) = \text{single-qb bitflip}, \\
        \dfrac13
        \left(1 - \dfrac{4p_i}{3} \right)^{-1/2}
        & \text{if } c(e) = \text{single-qb depolarizing}, \\
        \dfrac{1}{15}
        \left(1 - \dfrac{16p_i}{15} \right)^{-7/8}
        & \text{if } c(e) = \text{two-qb depolarizing}.
    \end{cases}
    \label{eq:fault_prob_derivative}
\end{equation}

\section{Variance Ratio}
\label{app:uncertainty}

For the differentiable estimator,
\begin{equation}
    \hat{\nu}_i^{\mathrm{Diff}}
    =
    \frac{1}{N}\sum_{k=1}^{N}S_{i,k},
    \qquad
    S_{i,k}
    =
    L(E_k)\pdv{p_i}\log \pr_{\pvec}(E_k)
    \equiv
    L(E_k)A_i(E_k).
\end{equation}
Using Eq.~\ref{eq:sens_model},
\begin{equation}
    A_i(E)
    =
    \sum_{e\in E}
    \pdv{W_e}{p_i}\frac{1}{W_e}
    -
    \sum_{e\notin E}
    \pdv{W_e}{p_i}\frac{1}{1-W_e}.
    \label{eq:appendix-score}
\end{equation}
Assuming independent shots,
\begin{equation}
    \mathrm{Var}\!\left(\hat{\nu}_i^{\mathrm{Diff}}\right)
    =
    \frac{\mathrm{Var}(S_i)}{N}.
    \label{eq:diff-var}
\end{equation}
For small \(p_L\),
\begin{align}
    \mathrm{Var}(S_i)
    &=
    \mathbb E[S_i^2]-\mathbb E[S_i]^2 \\
    &\approx
    \mathbb E[S_i^2] \\
    &=
    p_L\,\mathbb E[A_i^2\mid L].
    \label{eq:score-var-low-pl}
\end{align}

For small physical error rates, the DEM probabilities associated with channel \(i\) are linear to leading order,
\begin{equation}
    W_e(p_i)\approx \alpha_e p_i,
    \qquad
    \pdv{W_e}{p_i}\approx\alpha_e.
\end{equation}
Rearranging Eq.~\eqref{eq:appendix-score},
\begin{align}
    A_i(E)
    &=
    \sum_{e\in E}
    \pdv{W_e}{p_i}
    \left(\frac{1}{W_e}+\frac{1}{1-W_e}\right)
    -
    \sum_e
    \pdv{W_e}{p_i}\frac{1}{1-W_e} \\
    &=
    \frac{M_i(E)}{p_i}+O(1),
    \label{eq:Mi-scaling}
\end{align}
where \(M_i(E)\) is the effective number of channel-\(i\) faults in the sampled fault configuration. Hence
\begin{equation}
    \mathbb E[A_i^2\mid L]
    =
    \frac{\mathbb E[M_i^2\mid L]}{p_i^2}
    +O\!\left(\frac{1}{p_i}\right),
\end{equation}
and therefore
\begin{equation}
    \mathrm{Var}\!\left(\hat{\nu}_i^{\mathrm{Diff}}\right)
    \approx
    \frac{p_L}{Np_i^2}
    \mathbb E[M_i^2\mid L].
    \label{eq:diff-var-scaling}
\end{equation}

For central finite differences,
\begin{equation}
    \hat{\nu}_i^{\mathrm{FD}}
    =
    \frac{\hat p_L^+-\hat p_L^-}{2h},
    \qquad
    \hat p_L^{\pm}
    =
    \frac{1}{n_{\pm}}\sum_{k=1}^{n_{\pm}}L(E_k^{\pm}).
\end{equation}
Since \(\hat p_L^{\pm}\) are Bernoulli sample means,
\begin{equation}
    \mathrm{Var}(\hat p_L^{\pm})
    =
    \frac{p_L^{\pm}(1-p_L^{\pm})}{n_{\pm}}
    \approx
    \frac{p_L^{\pm}}{n_{\pm}},
\end{equation}
so
\begin{equation}
    \mathrm{Var}\!\left(\hat{\nu}_i^{\mathrm{FD}}\right)
    \approx
    \frac{p_L^+/n_+ + p_L^-/n_-}{4h^2}.
    \label{eq:fd-var}
\end{equation}

Let \(s\) be the number of error types. For equal total shot budget \(N\), finite differences use
\begin{equation}
    n_+=n_-=\frac{N}{2s}.
\end{equation}
Using \(p_L^++p_L^-\approx2p_L\),
\begin{equation}
    \mathrm{Var}\!\left(\hat{\nu}_i^{\mathrm{FD}}\right)
    \approx
    \frac{s p_L}{Nh^2}.
    \label{eq:fd-var-scaling}
\end{equation}
The variance ratio is therefore
\begin{align}
    R_i
    &\equiv
    \frac{\mathrm{Var}(\hat{\nu}_i^{\mathrm{FD}})}
    {\mathrm{Var}(\hat{\nu}_i^{\mathrm{Diff}})} \\
    &\approx
    \frac{s p_i^2}
    {h^2\,\mathbb E[M_i^2\mid L]}.
    \label{eq:ratio-general}
\end{align}
$ \expec[M_i^2\mid L]$ is the expected squared number of channel-$i$ faults involved in logical failures.

For a relative finite-difference step \(h=a p_i\),
\begin{equation}
    R_i
    \approx
    \frac{s}{a^2\,\mathbb E[M_i^2\mid L]}.
    \label{eq:ratio-relative-step}
\end{equation}
In our simulations, \(s=5\) and \(a=0.1\), giving
\begin{equation}
    R_i
    \approx
    \frac{500}{\mathbb E[M_i^2\mid L]}.
    \label{eq:ratio_exp}
\end{equation}

\begin{figure}
    \centering
    \includegraphics[width=0.8\linewidth]{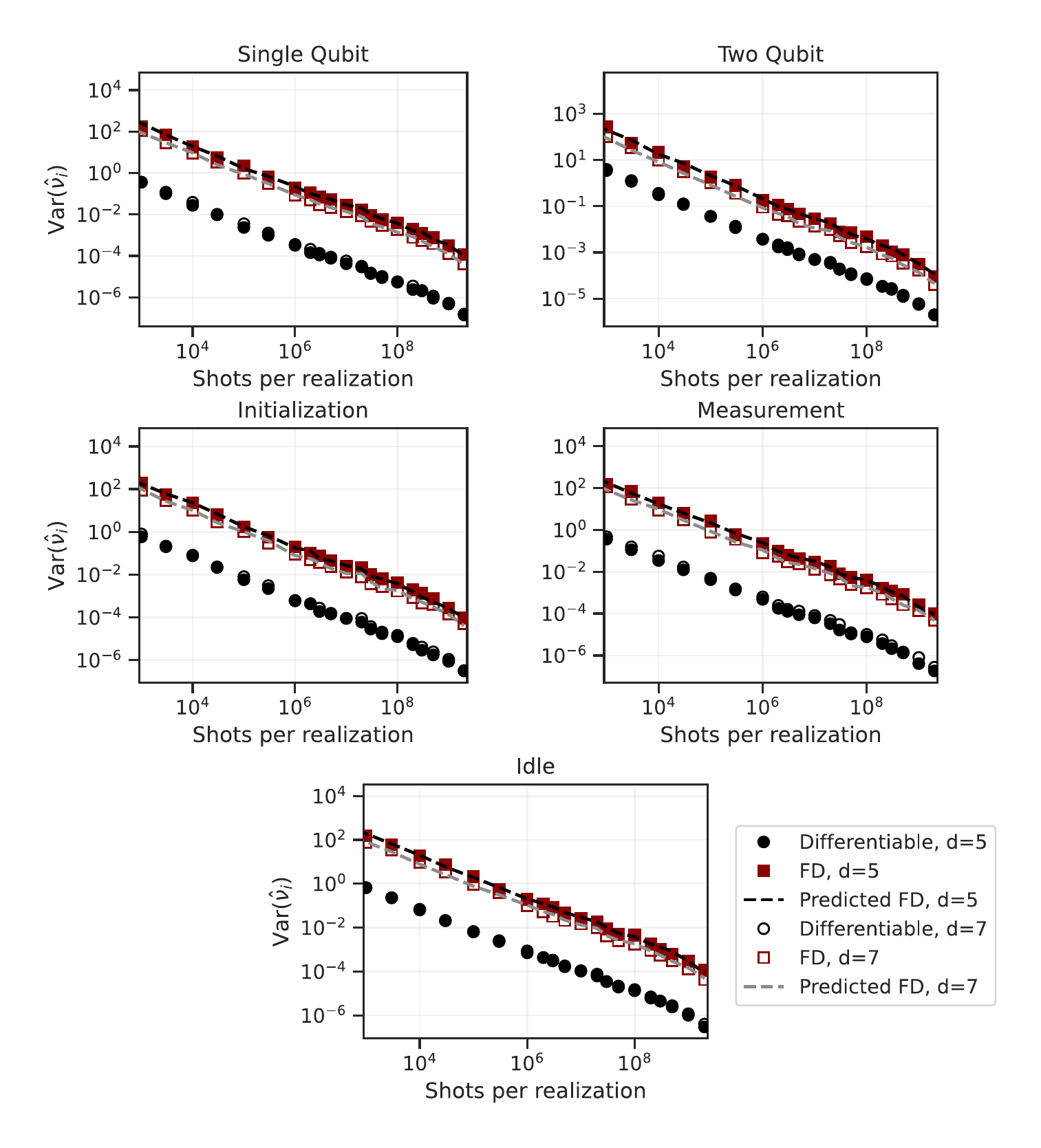}
    \caption{Comparing the variance for \fd{} predicted by $R_i$ with the measured variance.}
    \label{fig:variance_predicted}
\end{figure}

We used Eq.~\ref{eq:ratio_exp} to predict the value of $\mathrm{Var}(\hat \nu _i ^\text{FD})$ given $\expec[M_i^2 \mid L]$ and $\mathrm{Var}(\hat \nu _i ^\text{Diff})$, shown in Fig.~\ref{fig:variance_predicted}.
The predicted and measured values agree well, supporting Eq.~\ref{eq:ratio-general} as an approximation to the variance reduction achieved by \diff{} relative to \fd{}.

\end{document}